\documentstyle[11pt]{article}
\newlength{\bredde}
\def\slash#1{\settowidth{\bredde}{$#1$}\ifmmode\,\raisebox{.15ex}{/}
\hspace*{-\bredde} #1\else$\,\raisebox{.15ex}{/}\hspace*{-\bredde} #1$\fi}

\newcommand{\beq}{\begin{equation}}
\newcommand{\eeq}{\end{equation}}
\newcommand\beqn{\begin{eqnarray}}
\newcommand\eeqn{\end{eqnarray}}
\newcommand{\nn}{\nonumber}

\newcommand{\Tr}{\mbox{Tr}}
\newcommand{\del}{\partial}
\newcommand{\de}{\delta}
\newcommand{\th}{\theta}
\newcommand{\Pf}{{\cal P}_\phi}
\newcommand{\Pc}{{\cal P}}
\newcommand{\Pl}{{\cal P}_l}
\newcommand{\Zf}{{\cal Z}_\phi}
\newcommand{\Z}{{\cal Z}}
\newcommand{\Zl}{{\cal Z}_l}
\newcommand{\dM}{{\cal D}M}
\def\rar{\longrightarrow}
\def\l{\label}
\def\r{\ref}
\def\la{\lambda}
\def\Ga{\Gamma}
\def\eps{\epsilon}
\def\sig{\sigma}

\begin{document}
\topmargin -1.4cm
\oddsidemargin -0.8cm
\evensidemargin -0.8cm

\title{\Large{{\bf Non-universality of compact support probability 
distributions in random matrix theory}}}

\vspace{2.5cm}

\author{~\\{\sc G. Akemann $^{(1)}$, G.M. Cicuta $^{(2)}$, L. Molinari $^{(3)}$
 and G. Vernizzi $^{(2)}$ }\\~\\
 (1) Max-Planck-Institut f\"ur Kernphysik\\
Saupfercheckweg 1, D-69117 Heidelberg, Germany\\~\\
(2) Dipartimento di Fisica\\Viale delle Scienze, I-43100 Parma, Italy\\
and INFN Gruppo collegato di Parma\\~\\
(3) Dipartimento di Fisica and INFN\\Via Celoria 16, I-20133 Milano, Italy}

\date{}
\maketitle
\vfill
\begin{abstract}
The two-point resolvent is calculated in the large-$n$ 
limit for the generalized fixed and bounded trace ensembles. It is shown to 
disagree with the one of the canonical Gaussian ensemble by a non-universal 
part which is given explicitly for all monomial potentials $V(M)=M^{2p}$. 
Moreover, we prove that for the generalized fixed and bounded trace ensemble
all $k$-point resolvents agree in the large-$n$ limit, 
despite their non-universality.
\end{abstract}
\vfill

\newpage

\setcounter{equation}{0}
\section{Introduction}

Restricted Trace Ensembles (RTEs) introduced a long time ago in \cite{met}
are interesting for a couple of reasons. They possess compact 
support not only for infinite but also for finite $n$, where $n$ is the size 
of the matrix. In the canonical ensemble eq. (\r{Pcan})
large values of matrix elements are only exponentially suppressed whereas in
the RTEs a sharp cutoff is introduced. For this reason 
the latter can be regarded as the corresponding microcanonical ensemble. 

Much of the relevance of random matrix theory is related to universality 
properties
of connected correlators in the large $n$ limit, that is their independence 
from the
details of the probability density which defines the matrix ensemble. The most
famous property is the limiting form of connected density-density correlator at
``short distances'', also called the ``sine law''. Very interesting is also  a
global universality property: it was found that smoothed connected correlators
may be expressed by the same universal function. The original derivation
made use of loop equations \cite{AJM}, it was later rediscovered by 
diagrammatical
expansion \cite{BZ}. All derivations are valid for canonical ensembles with an
arbitrary polynomial, therefore it was generally believed that this global
universality property holds for all probability densities invariant under 
unitary transformations. In this note we investigate
two major  questions, namely whether the RTEs also possess universal 
global correlations, which are independent of the details of 
the distribution, and, second whether they are equivalent to the universality
classes of the 
canonical ensemble. Notice that usual techniques as orthogonal 
polynomials fail for RTEs because of 
the additional constraint on the matrix-trace.

In order to address to the above problems 
in a previous publication \cite{ACMV} 
we have introduced the following generalization of the RTEs 
\beqn
\Pf(M)&\equiv&\frac{1}{\Zf}\,\phi
\left( A^{2}-\frac{1}{n} \Tr  V(M)\right) \ \ \ \ \ \ , \ 
V(M)=\sum_{l=1}^p g_{2l}M^{2l} \ \ , \nn\\                              
\Zf &\equiv& \int \dM\,
 \phi \left(A^{2}-\frac{1}{n}\Tr V(M)\right) \ \ , 
\l{PRTE}
\eeqn
where $\phi(x)=\de(x)$ or $\th(x)$, and compared them to the canonical
ensemble
\beqn
\Pc(M)\equiv \frac{1}{\Z}\,\exp
\left[-ng\Tr V(M)\right] \ \ ,\ 
\Z \equiv \int \dM \,
 \exp \left[-ng\Tr V(M)\right] \ \ . 
\l{Pcan}
\eeqn
We have calculated the spectral density $\rho(\la)=<1/n\Tr\de(\la-M)>$ 
of RTEs, which is equivalent to the one-point resolvent 
$G(z)=<1/n\Tr (z-M)^{-1}>$. Comparing it to the canonical ensemble we 
have shown, that in the large-$n$ limit 
they agree  provided that the scale factor $g$ takes a well defined value 
determined by the values of the couplings $g_{2l}$
in the potential $V(M)$ and by $A^2$. This holds
despite the well known fact that the the spectral density itself is 
non-universal. From the factorization property of correlators at large-$n$
we then concluded that all finite moments of the three ensembles 
coincide. The question now is whether this equivalence holds also for the 
connected part of higher correlation functions and thus for higher orders in 
$1/n$. Therefore in this letter we investigate all $k$-point 
correlators. We start with $k=2$:
\beq
G_\phi (z,w) \equiv 
\left\langle\frac{1}{n}\Tr \frac{1}{z-M}
\frac{1}{n}\Tr\frac{1}{w-M}\right\rangle_\phi
\ - \ \left\langle\frac{1}{n}\Tr\frac{1}{z-M}\right\rangle_\phi
      \left\langle\frac{1}{n}\Tr\frac{1}{w-M}\right\rangle_\phi \ ,
\l{G2}
\eeq
where the subscript $\phi$ indicates the corresponding average. 
Here we have subtracted the factorized part. The 2-point correlator
as well as all higher $k$-point correlators are known to be 
universal for the canonical ensemble \cite{AJM}. There, the subtraction 
corresponds to taking into account only connected diagrams of random
surfaces. The corresponding connected density-density
correlators can be obtained by taking the appropriate
imaginary part, as given for example in \cite {AA}
\footnote{In contrast to ref. \cite{AA} we define the $k$-point resolvents 
without a factor of $n^{2k-2}$.}.

\setcounter{equation}{0}
\section{Non-universality of $G_\phi(z,w)$}

In the following we shall restrict ourselves to purely monomial 
potentials $V(M)=M^{2p}$. Since we want to show that the correlator
$G_\phi(z,w)$ is {\it non-universal}, in principle only two examples of 
different potentials leading to a different correlator eq. (\r{G2})
would be sufficient. In a first step we will show that all expectation
values of products of matrices have a $1/n^2$-expansion for the RTEs.
When relating the averages to the corresponding canonical ones we can
explicitly extract their expansion coefficients, which enables us to
calculate $G_\phi(z,w)$.

It is useful to introduce the following representation for the $\de$-
and $\th$-function 
\beq
\phi (x)=\int_{-\infty}^\infty \frac{dy}{2\pi }%
\frac{e^{(iy+\eps)x}}{(iy+\eps)^s} \ \ , \left\{ \begin{array}{ll}
                                                 s=0 & \phi(x)=\de(x)\\
                                                 s=1 & \phi(x)=\th(x)
                                                 \end{array}
                                         \right.\ ,
\l{phi}
\eeq                    
where $\eps=0^{+}$ is a small and harmless
regulator which makes possible to interchange integrals.
Next we calculate the matrix integral
\beq
I^{\{k\}}_\phi(n,A)\equiv\int {\cal{D}}M\,\phi \left( A^{2}-\frac{1}{n}
\Tr M^{2p}\right) \;M_{\alpha _{1}\beta _{1}}M_{\alpha _{2}\beta _{2}}
\ldots M_{\alpha _{k}\beta _{k}}     \ \ ,
\l{Idef}
\eeq
where the superscript $\{k\}$ summarizes the dependence on all the matrix 
indices.
The volume element of the Hermitean $(n\times n)$-matrices 
$\prod_i  dM_{ii} \prod_{i<j} d \mbox{Re} M_{ij} \prod_{i<j} d \mbox{Im} 
M_{ij}$ is the usual product of independent entries.
In the particular case $k=0$ eq. (\r{Idef}) is just the 
partition function $\Zf$.
By inserting the representation eq. (\r{phi}) into the eq. (\r{Idef}) and 
exchanging the order of integrations, we  exhibit that (\r{Idef}) is actually 
proportional to the analogous integral with canonical measure. Indeed
\beq
I_\phi^{\{k\}}(n,A) =
\int_{-\infty }^\infty\frac{dy}{2\pi} e^{(iy+\eps)A^{2}}
\frac{1}{(iy+\eps)^s}  
\int {\cal{D}}M\;e^{- [(iy+\eps)/n]  \Tr M^{2p}}\;M_{\alpha _{1}\beta_{1}}
\ldots M_{\alpha _{k}\beta _{k}}  \ .
\label{2.1}
\eeq
 The explicit dependence of the matrix integral
\beq
\int {\cal{D}}M\;e^{- a \, \Tr M^{2p}}\;M_{\alpha _{1}\beta_{1}}
\ldots M_{\alpha _{k}\beta _{k}}  =
a^{-\frac{n^2+k}{2p} }
\int {\cal{D}}M\;e^{- \Tr M^{2p}}\;M_{\alpha _{1}\beta_{1}}
\ldots M_{\alpha _{k}\beta _{k}}  
\label{2.2}
\eeq
on the real positive parameter $a$, allows analytic continuation in the whole
complex plane, cut along the negative part of the real axis. This provides a
definition for (\r{2.1}) and we obtain
\beqn
I_\phi^{\{k\}}(n,A) &=& 
\int_{-\infty }^\infty\frac{dy}{2\pi} e^{(iy+\eps)A^{2}}
\frac{1}{(iy+\eps)^s}  
\left( \frac{gn^2}{iy+\eps}\right) ^{\frac{n^{2}+k}{2p}}
\int {\cal{D}}M\;e^{-ng\Tr M^{2p}}\;M_{\alpha _{1}\beta_{1}}
\ldots M_{\alpha _{k}\beta _{k}}  
\nn\\
&=& \left(gn^2A^{2}\right)^{\frac{n^{2}+k}{2p}}
\frac{\left( A^2\right)^{s-1}}{\Gamma \left( \frac{n^{2}+k}{2p}+s\right)}\;
\int {\cal{D}}M\;e^{-ng\Tr M^{2p}}\;M_{\alpha _{1}\beta_{1}}
\ldots M_{\alpha _{k}\beta _{k}}\ \ .
\l{I}
\eeqn
where $g>0$ and in the second step we have used Hankel's contour integral 
for the Gamma 
function \cite{Abramo}. As a consequence we obtain the RTE average 
expressed by the canonical average
\beq
\left\langle M_{\alpha _{1}\beta _{1}}\ldots
M_{\alpha _{k}\beta _{k}}\right\rangle _\phi 
\ =\ \frac{I_\phi^{\{k\}}(n,A)}{I_\phi^{\{0\}}(n,A)} 
\ =\ s_{n,k}(g,A) \left\langle M_{\alpha _{1}\beta _{1}}\ldots
M_{\alpha _{k}\beta _{k}}\right\rangle \ ,                  \label{mom}
\eeq
where 
\beq
s_{n,k}(g,A) \ = \ \left( gn^2A^2\right) ^\frac{k}{2p}\frac{\Gamma \left( 
\frac{n^{2}}{2p}+s\right) }{\Gamma \left( \frac{n^{2}+k}{2p}+s\right)} \ .
\l{sn}
\eeq
On the r.h.s. of eq. (\r{mom}) we average with the canonical measure 
eq. (\r{Pcan}) for $V(M)=gM^{2p}$.  The exact relation (\r{mom})
may be exploited to relate the parameters of the RTEs
to the parameters of the canonical model, so that at leading order in the
large-$n$ limit all moments of the  form  (\r{mom}) are identical. It is 
however impossible to relate the parameters to obtain 
that the scaling factor $s_{n,k}(g,A)$ is unity up to order
$O(1/n^4)$. Indeed, if we assume 
\beq
A^2 = \frac{1}{2 p g } + \frac{x}{n^2} + O\left(\frac{1}{n^4}\right)
\l{2.3}
\eeq
and use the relation for ratios of Gamma functions \cite{Abramo}, we obtain
\beqn
s_{n,k}(g,A) &=& (2p\ g A^2)^\frac{k}{2p}
\left[ 1 - \frac{k}{2 n^2}\left( \frac{k}{2p}+2s-1\right)
 +  O\left(\frac{1}{n^4}\right)\right] 
\nonumber \\
&=&  1+\frac{k}{n^2}\left( g x - \frac{k}{4 p} -s +\frac{1}{2} \right) +
 \ O\left(\frac{1}{n^4}\right) \ \ .
\l{sexp} 
\eeqn
This shows that, with the general relation (\r{2.3}), the  $1/n^2$ expansion
of the canonical measure translates into a  $1/n^2$ expansion for the
RTEs \footnote{Using Stirling's formula one can easily
convince oneself that $z^{b-a}\frac{\Ga(z+a)}{\Ga(z+b)}$ has an expansion in
$1/z$.} . The non-vanishing contribution at order $1/n^2$ in the scaling factor
$s_{n,k}(g,A)$ with the   $k^2$-dependence will be shown
 to lead to the non-universality of connected correlators. 

The relation between the coefficients $c_{\{k\}}^{(j)}$
of the topological $1/n^2$-expansion in the canonical ensemble are simply
related to the corresponding coefficients $d_{\{k\}}^{(j)}$ for the RTEs
\beq
\left\langle M_{\alpha _{1}\beta _{1}}\ldots
M_{\alpha _{k}\beta _{k}}\right\rangle \ =\ \sum_{j=0}^\infty c_{\{k\}}^{(j)}
\frac{1}{n^{2j}} \ \ , \ \
\left\langle M_{\alpha _{1}\beta _{1}}\ldots
M_{\alpha _{k}\beta _{k}}\right\rangle_\phi 
\ =\ \sum_{j=0}^\infty d_{\{k\}}^{(j)}\frac{1}{n^{2j}} 
\l{ddef}
\eeq
through eq. (\r{sexp})
\beqn
d_{\{k\}}^{(0)} &=& c_{\{k\}}^{(0)} \ \ ,\nn\\
d_{\{k\}}^{(1)} &=& c_{\{k\}}^{(1)} 
    +k \left( g x -\frac{k}{4p}-s+\frac{1}{2}\right)c_{\{k\}}^{(0)} \ \ ,
\l{dc}
\eeqn
where we recall that we have $s=0,1$ for $\phi=\de,\th$ and the subscript
$\{k\}$ summarizes all matrix indices. 
Eq. (\r{dc}) immediately implies the identity of the one-point resolvents
\cite{ACMV} 
\beqn
G_\phi(z) &=& \frac{1}{n}\sum_{k=0}^\infty\frac{1}{z^{k+1}}
\left\langle\Tr M^k\right\rangle_\phi 
\ = \ \frac{1}{n}\sum_{k=0}^\infty\frac{1}{z^{k+1}}\left(d_{\{k\}}^{(0)}
+O\left(\frac{1}{n^2}\right)\right) \nn\\
&\stackrel{n\to\infty}{\rar}& G(z) \ \ ,
\l{GG1}
\eeqn
which has been shown in \cite{ACMV} for a larger class
of potentials. Note that $G_\phi(z)$ 
is of order 1 since $d_{\{k\}}^{(0)}$ contains
a power of $n$ from the trace. 

Next we turn to the two-point resolvent $G_\phi(z,w)$. Inserting eq. (\r{dc})
into the definition (\r{G2}) we obtain
\beqn
G_\phi(z,w) 
&=&\frac{1}{n^{2}}\sum_{k,l=0}^{\infty }\frac{1}{z^{k+1}w^{l+1}}
\left(\left\langle \Tr M^{k}\Tr M^{l}\right\rangle _\phi
-\left\langle \Tr M^{k}\right\rangle _\phi\left\langle 
\Tr M^{l}\right\rangle _\phi\right)                     \nn\\
&=&\frac{1}{n^{2}}\sum_{k,l=0}^{\infty }\frac{1}{z^{k+1}w^{l+1}}
\left( d_{\{ k,l\} }^{(0)}+d_{\{ k,l\} }^{(1)}\frac{1}{n^{2}}
- \left[ d_{\{ k\} }^{(0)}+d_{\{ k\} }^{(1)}\frac{1}{n^{2}}\right] 
\left[ d_{\{ l\}}^{(0)}+d_{\{ l\} }^{(1)}\frac{1}{n^{2}}\right]   
+ O\left(\frac{1}{n^{4}}\right)\right) \nn\\
&=&\frac{1}{n^{4}}\sum_{k,l=0}^{\infty }\frac{1}{z^{k+1}w^{l+1}}
\left( c_{\{ k,l\} }^{(1)}
- c_{\{ k\}}^{(0)}c_{\{ l\} }^{(1)} - c_{\{ k\}}^{(1)}c_{\{ l\} }^{(0)}
- \frac{kl}{2p}c_{\{ k\}}^{(0)}c_{\{ l\} }^{(0)}
+ O\left(\frac{1}{n^{2}}\right)\right)\ .
\l{GG2a}
\eeqn
Here we have made use of the fact that
\beqn
d_{\{ k,l\} }^{(0)} &=&    c_{\{ k,l\} }^{(0)}  =      
         c_{\{k\} }^{(0)}c_{\{ l\} }^{(0)} \ ,
\nonumber \\  
d_{\{ k,l\} }^{(1)} &=& c_{\{ k,l\} }^{(1)} 
+(k+l)\left( g x -\frac{k+l}{4p} -s +\frac{1}{2}
\right) c_{\{ k,l\} }^{(0)} \ \ .
\l{dcc}
\eeqn
 As a consequence the leading
terms in $G_\phi(z,w)$ cancel as they should, 
leaving the remaining part of order
$1/n^2$  when counting properly factors of $n$ from the traces.
Remarkably the result (\r{GG2a}) does not depend upon the values of $x$ 
and $s$.
The first three terms in the last line of eq. (\r{GG2a}) give precisely
the universal two-point resolvent of the canonical ensemble. The last term
is new and can be written as a product of derivatives of the one-point
resolvent $G(z)$ $(=G_\phi(z))$. The final result reads
\beq
n^2G_\phi(z,w)\ \stackrel{n\to\infty}{\rar}\ n^2G(z,w) \ -\ 
\frac{1}{2p}\del_z(zG(z))\del_w(wG(w)) \ \ ,
\l{GG2} 
\eeq
which holds for all monomial potentials $V(M)=gM^{2p}$ ,
both $\phi=\de,\th$ and  $A^2$ given by eq.(\r{2.3}).
Note that all terms in eq. (\r{GG2}) are of order 1.
The first term is the well known universal two-point resolvent \cite{AJM}
\beqn
n^2G(z,w) &=& \frac{1}{2(z-w)^2}\left(
\frac{zw \ -\ a^2}{\sqrt{(z^2-a^2)(w^2-a^2)}} \ -\ 1\right) \ ,
\l{Guniv}
\eeqn
where $a$ denotes the support of the eigenvalues $[-a,\ a]$.
The notion of universality means that
eq. (\r{Guniv}) is the same for any given polynomial potential sharing
the same support \cite{AJM}. We only have to assume that the couplings 
$g_{2l}$ are such that the support is one arc. This is true in particular
for the monomial potentials with $g>0$.
The second term in eq. (\r{GG2}), however, is {\it non-universal} as the
one-point resolvent itself is non-universal. Let us give two examples, the 
Gaussian and the purely quartic potential:
\beqn
\underline{V(M)=gM^2}:\ \ \ G(z)&=&g\left(z - \sqrt{z^2-a^2}\right)
\ \ \ \ \ \ \ \ \ \ \ \ \ \ \ \ \ \ \ , \ a^2 = \frac{2}{g} \ ,\nn\\ 
\underline{V(M)=gM^4}:\ \ \ G(z)&=&g
                          \left(2z^3 - (2z^2+a^2)\sqrt{z^2-a^2}\right)
\ \ , \ a^4 = \frac{4}{3g} \ .
\l{Gexpl}
\eeqn
Although in this case we have a potential depending only on one parameter $g$,
which is thus in one to one correspondence to the endpoint of support $a$,
the two resolvents in eq. (\r{Gexpl}) are different {\it functions} of $z$.
Inserting them into the result for $G_\phi(z,w)$ eq. (\r{GG2}) we obtain
\beqn
\underline{V(M)=gM^2}&:& \nn\\ 
n^2G_\phi(z,w)&=& n^2G(z,w) - g^2 
\left(2z-\frac{2z^2-a^2}{\sqrt{z^2-a^2}}\right)
\left(2w-\frac{2w^2-a^2}{\sqrt{w^2-a^2}}\right) \ ,
\nn\\
\underline{V(M)=gM^4}&:& \nn\\ 
n^2G_\phi(z,w)&=& n^2G(z,w) - g^2 
\left(8z^3-\frac{8z^4-4a^2z^2-a^4}{\sqrt{z^2-a^2}}\right)
\left(8w^3-\frac{8w^4-4a^2w^2-a^4}{\sqrt{w^2-a^2}}\right) \ , \nn\\
\l{G2non}
\eeqn
where $n^2G(z,w)$ is given in eq. (\r{Guniv}) and $A^2$ in eq.(\r{2.3}).
These examples clearly demonstrate the non-universality of
$G_\phi(z,w)$, which cannot be repaired by a suitable parameter redefinition.

Let us finally mention that we have verified eq. (\r{GG2}) for the quadratic
potential following an entirely different approach. As it has been already
emphasized in \cite{ACMV} the fixed trace ensemble $\phi=\de$ can be obtained
from the ``trace squared ensemble''
\beqn
\Pl(M)&\equiv& \frac{1}{\Zl}\,\exp
\left[-l\left( -2nA^2\Tr V(M)  + (\Tr V(M))^2\right)\right] \ \ , \ \nn\\
\Zl &\equiv& \int \dM \,
 \exp \left[-l\left( -2nA^2\Tr V(M)  + (\Tr V(M))^2\right)\right] \ \ , 
\l{Psq}
\eeqn
when taking the limit $l\to\infty$. The trace square terms add so-called 
``touching'' interactions to the triangulated surface \cite{das}.
This representation of the fixed trace ensemble does not only provide us
with a different technical tool to check eq. (\r{GG2}) for $p=1$, which we do 
not display here.
It also gives us a diagrammatical interpretation in terms of Feynman graphs,
which explains the existence of the $1/n^2$-expansion in eq. (\r{ddef})
for the Gaussian fixed trace ensemble as a topological expansion.

It seems remarkable that the second term in eq. (\r{GG2}), in the case of
Gaussian resolvent $G_\phi(z) \to G(z) =\frac{2}{a^2} (z- \sqrt{z^2-a^2})$ ,
has the same form of the analogous term which appears in the connected
correlator for Wigner ensembles \cite{pastur}, \cite{danna}, 
written in different but equivalent forms since
\beqn
\del_z\left( z G(z) \right)=
\frac{-2 a^2 [G(z)]^3}{4-a^2  [G(z)]^2}
=\frac{a^2}{4} \del_z[G(z)]^2 \ \ .
\nonumber
\eeqn

\setcounter{equation}{0}
\section{Equivalence of all higher-point resolvents of the RTEs}

The two-point resolvent of the fixed and bounded trace ensemble has turned 
out to be identical in the large-$n$ limit although non-universal. 
It is therefore natural to ask whether this equivalence holds also for
all higher $k$-point resolvents. In the following we will show that this is 
indeed the case.
Let us define the two generating functionals
\beqn
Z_\phi[J] &\equiv& \sum_{k=0}^\infty \frac{1}{k!}\int dz_1\ldots dz_k
\left\langle\frac{1}{n}\Tr\frac{1}{z_1-M}\ldots
\frac{1}{n}\Tr\frac{1}{z_k-M}\right\rangle_\phi J(z_1)\ldots J(z_k) \ \ ,
\l{ZJ}\\
W_\phi[J] &\equiv& \sum_{k=0}^\infty \frac{1}{k!}\int dz_1\ldots dz_k
\ G_\phi(z_1,\ldots,z_k)\ J(z_1)\ldots J(z_k)  \ \ ,
\l{WJ}
\eeqn
where 
\beq
G_\phi(z_1,\ldots,z_k) \equiv 
\left\langle\frac{1}{n}\Tr\frac{1}{z_1-M}\ldots
\frac{1}{n}\Tr\frac{1}{z_k-M}\right\rangle_\phi 
- \sum_{\sig\in P} G_\phi(z_{\sig(1)},..,z_{\sig(l_1)}) \ldots 
G_\phi(z_{\sig(l_{k-1}+1)},..,z_{\sig(k)}) 
\l{Gk}
\eeq
is the $k$-point resolvent. The sum runs over all different partitions $P$
of the $k$ arguments and thus over all different combinations of 1- to $(k-1)$-
point resolvents with in total $k$ indices.
In the canonical ensemble this subtraction corresponds to taking only 
connected graphs into account. For this reason the $k$-point resolvent
is there of the order $1/n^{2k-2}$.
From field theory we know that the following relation between the 
generating functionals holds:
\beq
Z_\phi[J] \ =\ e^{W_\phi[J]} \ \ .
\l{ZW}
\eeq
The correlators can be obtained in the usual way
\beq
\frac{\de^k}{\de J^k} \left\{ \begin{array}{l}Z_\phi[J]\\W_\phi[J]\end{array}
\right|_{J=0} \ =\ \left\{
\begin{array}{l}\langle\frac{1}{n}\Tr\frac{1}{z_1-M}\ldots
\frac{1}{n}\Tr\frac{1}{z_k-M}\rangle_\phi \\
G_\phi(z_1,\ldots,z_k)
\end{array}\right. \ .
\eeq
In \cite{ACMV} it has been shown that the ensemble averages of the fixed and
bounded trace ensemble can be related:
\beq
\left\langle\ {\cal O}(M)\ \right\rangle_\de\ =\ 
(1+c_n\del_{A^2})\left\langle\ {\cal O}(M)\ \right\rangle_\th \ \ ,
\l{rel}
\eeq
where we have 
\beq
c_n \ =\ 2pA^2\frac{1}{n^2} \ \ \mbox{for} \ \ V(M)=M^{2p} \ .
\l{cn}
\eeq
Consequently the same relation holds for the generating functionals
$Z_\phi[J]$ following their definition (\r{ZJ})
\beq
Z_\de[J] \ =\ (1+c_n\del_{A^2})Z_\th[J] \ \ .
\l{ZZ}
\eeq 
Using the relation eq. (\r{ZW}) we can translate this to the 
generating functional for the resolvent operators
\beq
e^{W_\de[J]} \ =\ \left( 1+ c_n (\del_{A^2}W_\th[J])\right)
e^{W_\th[J]} \ \ ,
\eeq
or equivalently
\beq
W_\de[J]\ =\ W_\th[J]+\sum_{l=1}^\infty (-)^{l+1}\frac{1}{l}
(c_n\del_{A^2}W_\th[J])^l \ \ ,
\eeq
where we have expanded the logarithm. Taking the functional derivative
$\de^k/\de J^k$ and setting $J=0$ will truncate the infinite sum for the
following reason. From the definition we have $W_\phi[J\!=\!0]=1$ and thus
$\del_{A^2}W_\phi[J\!=\!0]=0$. For 
this reason only terms will persist
where at least one functional derivative $\de/\de J$ acts on 
$\del_{A^2}W_\phi[J]$. We finally obtain
\beqn
G_\de(z_1,\ldots,z_k) &=& (1+c_n\del_{A^2})G_\th(z_1,\ldots,z_k) \nn\\
 &+& \sum_{\sig\in P;\ l=2,..,k} (-)^{l+1}\frac{1}{l}
\left(c_n\del_{A^2}G_\th(z_{\sig(1)},..,z_{\sig(l_1)})\right)\ldots
\left(c_n\del_{A^2}G_\th(z_{\sig(l_{k-1}+1)},..,z_{\sig(k)})\right) .\nn\\
\l{GGk}
\eeqn
Here the sum runs again over all partitions $P$ of the $k$ arguments and
$l$ counts the number of blocks or resolvents 
into which the arguments are divided.
To prove the desired equivalence between the two $k$-point resolvents
we need to know the order in $1/n^2$ of all terms on the r.h.s. 
Following the diagrammatic approach mentioned at the end of the 
last section, where the fixed trace ensemble is represented by the trace
squared one, we obtain the same counting of powers as in the canonical 
ensemble already mentioned 
\beq
G_\de(z_1,\ldots,z_k)\ =\ O\left(\frac{1}{n^{2k-2}}\right) \ ,
\l{GOn}
\eeq
at least in the Gaussian case. In the following we will assume that same 
holds for the monomial potentials. We have checked this explicitly for
the 2- and 3-point resolvent using the definition (\r{Gk}) and the relation
(\r{dc}). It now follows easily by induction that eq. (\r{GOn}) also holds
for the bounded trace ensemble and that we have
\beq
n^{2k-2}G_\th(z_1,\ldots,z_k)\ \stackrel{n\to\infty}{\rar}\ 
n^{2k-2}G_\de(z_z,\ldots,z_k) \ , 
\l{equi}
\eeq
which generalizes eq. (\r{GG2a}) for the two-point resolvents.
Namely in eq. (\r{GGk}) on the r.h.s. the second term in the first
line is obviously subleading, due to $c_n\sim1/n^2$. Using induction 
in the sum each term is of the order 
$O(n^{-(2l_1+2(l_2-l_1)+..+2(k-l_{k-1})})$ = $O(n^{-2k})$ 
which is also subleading.

In the above derivation no explicit use has been made of the $\de$- or 
$\th$-measure apart from the fact that $\th^{\prime}(x)=\de(x)$. Instead of 
this we could have used for example $\phi(x)=x\th(x)$ and 
$\phi(x)=\th(x)$ because of
$(x\th(x))^{\prime}=\th(x)$. More generally we can extend the proof of relation
(\r{equi}) to an infinite class of RTEs with
\beq
\phi(x)=\left\{ \ \de(x),\ \th(x),\  
\left(\frac{1}{j!}x^j\th(x)\right)_{j\in \mbox{N}_+}\right\}\ \ , 
\l{newphi}
\eeq
showing that all
their $k$-point resolvents are equivalent at large-$n$. We only need to show
the starting point for $k=2$ since we have used induction. This can be shown
as follows. When we calculate the matrix integral 
$I^{\{k\}}_\phi(n,A)$ in eq. (\r{Idef}) we allow the parameter $s$ in
the representation eq. (\r{phi}) to take all non-negative integer values,
which is then a representation for all the measures introduced in 
eq. (\r{newphi}). The same derivation goes through up to the result for the
two-point resolvent eq. (\r{GG2}) as we have kept $s$ general and explicit
everywhere.

Let us conclude this section with a final remark. In ref. \cite{ACKM}
a topological expansion was introduced and calculated for each resolvent
\beq
G(z_1,\ldots,z_k) \ =\ \sum_{h=0}^\infty \frac{1}{n^{2h}} G_h(z_1,\ldots,z_k) 
\ .
\eeq
If we introduce the same expansion here for the $G_\phi(z_1,\ldots,z_k)$,
a short look at relation (\r{GGk}) tells us that already for $h=1$
(``genus one'') the equivalence eq. (\r{equi}) breaks down:
\beq
n^{2k-2+2h}G_{h\geq1,\th}(z_1,\ldots,z_k) \ \neq\ 
n^{2k-2+2h}G_{h\geq1,\de}(z_1,\ldots,z_k) \ .
\eeq
In this sense we have shown that only the ``planar'' $(h=0)$ $k$-point 
resolvents of the fixed and extended bounded trace ensembles agree.

\setcounter{equation}{0}
\section{Conclusions}

We have proved the non-universality of the two-point resolvent $G_\phi(z,w)$ 
of the generalized fixed and bounded trace ensembles by comparing it to the 
universal two-point resolvent of the canonical ensemble. 
Apart from the general results for $G_\phi(z,w)$ for all monomial potentials
$V(M)=M^{2p}$ we have explicitly displayed its non-universal parts 
in two examples, the quadratic and the pure quartic potential.

Furthermore, we have extended the equivalence of the generalized fixed
and generalized bounded ensemble in the large-$n$ limit from all finite 
moments \cite{ACMV} to all $k$-point resolvents, which probe higher
orders in $1/n^2$.

While we have shown that global universality
fails for the generalized RTEs, the issue of universality of correlators
at short distance, possibly  matching with the canonical ensemble,
is still open. We plan to come back to this interesting question in 
the future.

\end{document}